\documentclass[english,aps,prl,twocolumn,amsmath,amssymb,showpacs,superscriptaddress,notitlepage,longbibliography]{revtex4-2}
\usepackage{graphicx}% Include figure files
\usepackage{dcolumn}% Align Table columns on decimal point
\usepackage{bm}% bold math
\usepackage[usenames,dvipsnames]{color}
\usepackage[most]{tcolorbox}
\usepackage{multirow}
\usepackage{gensymb}
\usepackage{amssymb}
\usepackage{algorithm}
\usepackage{algorithmic}
\usepackage[normalem]{ulem}
\usepackage{CJK}
\usepackage{comment}
\usepackage{amsfonts}
\usepackage[colorlinks, linkcolor=blue,anchorcolor=blue,citecolor=blue,urlcolor=blue]{hyperref}
\usepackage{amssymb}
\usepackage{pifont}
\usepackage{physics}
\usepackage{amsmath}
\usepackage{natbib}
\usepackage{xcolor}
\usepackage{color,soul}

\begin{document}

%\title{Quantum Geometric Planar Magnetotransport: A Probe for Magnetic Geometry in Altermagnets}

\title{Quantum-Geometric Fingerprints of Altermagnetic Order in Planar Magnetotransport}

\author{Zhi-Chun Ouyang}
\thanks{These authors contributed equally.}
\affiliation{Department of Physics, The Hong Kong University of Science and Technology, Clear Water Bay, Hong Kong SAR, China}
   
\affiliation{Center for Theoretical Condensed Matter Physics, The Hong Kong University of Science and Technology, Clear Water Bay, Hong Kong SAR, China}
\author{Wei-Jing Dai}
\thanks{These authors contributed equally.}
\affiliation{Department of Physics, The Hong Kong University of Science and Technology, Clear Water Bay, Hong Kong SAR, China}
   
\affiliation{Center for Theoretical Condensed Matter Physics, The Hong Kong University of Science and Technology, Clear Water Bay, Hong Kong SAR, China}

\author{Zi-Ting Sun}\thanks{zsunaw@connect.ust.hk}

\affiliation{Department of Physics, The Hong Kong University of Science and Technology, Clear Water Bay, Hong Kong SAR, China}
   
\affiliation{Center for Theoretical Condensed Matter Physics, The Hong Kong University of Science and Technology, Clear Water Bay, Hong Kong SAR, China}

\author{Jin-Xin Hu}\thanks{jhuphy@ust.hk}
\affiliation{Department of Physics, The Hong Kong University of Science and Technology, Clear Water Bay, Hong Kong SAR, China}
   
\affiliation{Center for Theoretical Condensed Matter Physics, The Hong Kong University of Science and Technology, Clear Water Bay, Hong Kong SAR, China}
   
\author{K. T. Law }
\affiliation{Department of Physics, The Hong Kong University of Science and Technology, Clear Water Bay, Hong Kong SAR, China}
   
\affiliation{Center for Theoretical Condensed Matter Physics, The Hong Kong University of Science and Technology, Clear Water Bay, Hong Kong SAR, China}

\begin{abstract}
Identifying altermagnetic order through transport requires signatures that are sensitive to magnetic symmetry but do not rely on a net magnetization. Here we show that planar magnetotransport provides such quantum-geometric fingerprints. In two-dimensional altermagnets with $C_n\mathcal{T}$ magnetic symmetry, an in-plane Zeeman field explicitly breaks the mirror and emergent $C_{2z}$ symmetries that otherwise suppress intrinsic Hall and second-order transport responses. The resulting magnetic field susceptibilities of the Berry curvature and quantum metric produce linear planar Hall, nonlinear planar Hall, and nonreciprocal longitudinal responses. Crucially, the leading magnetic field powers and angular periodicities of these responses are fixed by the underlying altermagnetic order. For $d$-, $g$-, and $i$-wave altermagnets, we find distinct fingerprint patterns associated with quantum geometric susceptibilities. Our results establish planar magnetotransport as a symmetry selective probe of both band quantum geometry and altermagnetic order.

\end{abstract}

\maketitle

\emph{Introduction.}---Altermagnets have recently emerged as a distinct class of collinear magnetic materials that combine zero net magnetization with momentum dependent spin splitting~\cite{PhysRevX.12.031042,hayami2019momentum,Ma2021,Zhu2024,Zhu2025,ds5z-fxy5,cjzw-j4v7,PhysRevLett.134.136301,zm5y-vy41,Jungwirth2026,doi:10.1021/acs.nanolett.6c00136,PhysRevLett.133.206702,PhysRevLett.134.106802,PhysRevLett.134.166701,Jungwirth2025,https://doi.org/10.1002/adfm.202409327, Song2025, Peng2025,zt4l-y18j,Yang2026}. Unlike conventional antiferromagnets, where magnetic sublattices are connected by translation or inversion, altermagnets are characterized by spin space and real space operations that give rise to even-parity spin splitting in momentum space. In the absence of spin-orbit coupling (SOC), this structure can be described by a spin-group operation combining spin reversal and a $C_n$ spatial rotation; in the presence of SOC, it is encoded by a $C_n\mathcal{T}$ magnetic symmetry~\cite{PhysRevX.12.040501,PhysRevX.12.031042,Zhu2025}. We refer to this symmetry-imposed angular structure of the spin texture as the magnetic geometry of an altermagnet. A central challenge is to identify transport signatures that distinguish different altermagnetic orders, such as $d$-, $g$-, and $i$-wave orders, without relying on a net magnetization.

Quantum-geometric transport offers a natural route towards this problem. Nonlinear Hall effects, nonreciprocal magnetochiral anisotropy, and nonlinear optical responses are known to encode Berry curvature, Berry curvature multipoles, and quantum-metric effects within Bloch bands~\cite{PhysRevX.14.021046,PhysRevLett.130.266003,doi:10.1126/science.adq3255,PhysRevB.111.174513,PhysRevB.106.L041111,PhysRevLett.134.026001,PhysRevB.109.L201403,Tzschaschel2024, Wu2019,PhysRevB.108.L201405,PhysRevLett.115.216806,PhysRevB.107.115142, Lu2024,PhysRevLett.117.146603,Tokura2018,PhysRevLett.132.026002, Hu2025,PhysRevResearch.7.023273,PhysRevLett.132.196801,l3c7-knqm,zt4l-y18j,Ezawa_2026,liang2026kerrlikeeffectinducedquantummetric}. Recent observations of nonlinear Hall and nonreciprocal transport in $\mathrm{MnBi}_2\mathrm{Te}_4$ further demonstrate the sensitivity of such responses to unconventional magnetic states~\cite{Wang2023,PhysRevLett.132.026301,Li2024,PhysRevLett.127.277201,PhysRevLett.127.277202}. In altermagnets, however, the same symmetries that define the magnetic order can also hide the most direct quantum-geometric responses: mirror symmetry suppresses the linear anomalous Hall effect, while an emergent $C_{2z}$ symmetry generally forbids intrinsic second-order transport~\cite{PhysRevB.111.184407,PhysRevLett.133.106701,rv1n-vr4p}. The key problem is therefore not merely to detect a transport response, but to construct quantum-geometric fingerprints that are symmetry selective for underlying altermagnetic orders.

\begin{figure}
    \centering
    \includegraphics[width=1\linewidth]{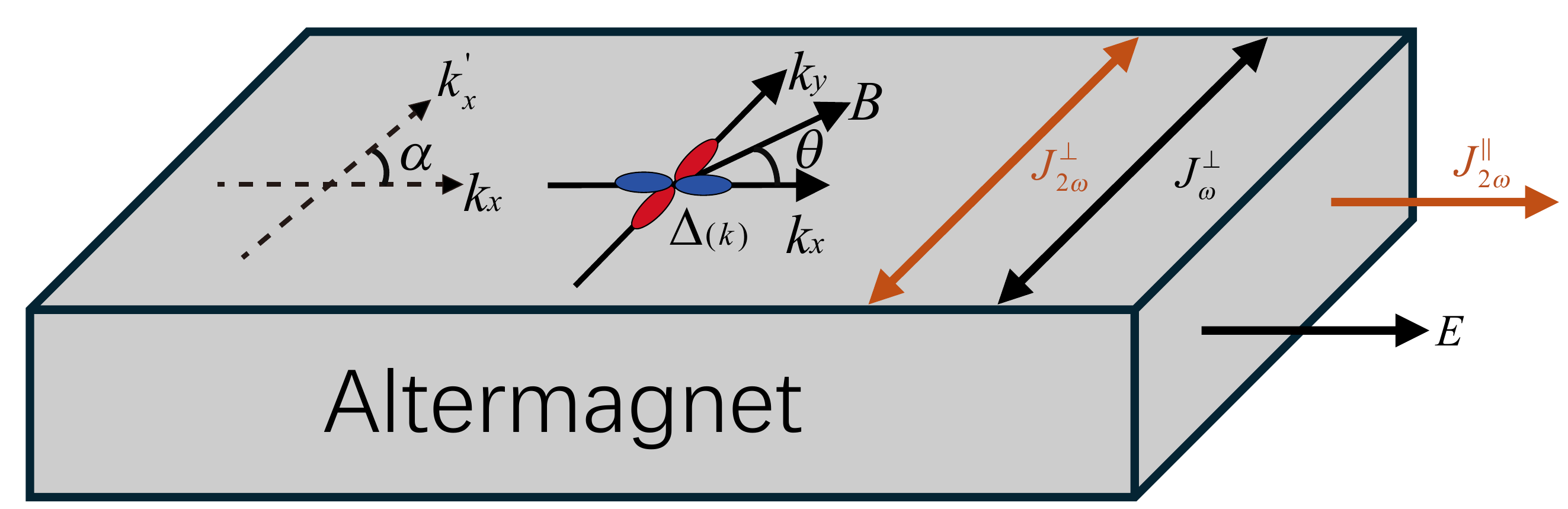}
    \caption{Schematic of planar magnetotransport fingerprints in altermagnets with $C_n \mathcal{T}$ symmetry. The in-plane magnetic field breaks the relevant mirror and $C_{2z}$ constraints through Zeeman coupling, generating the linear planar Hall response $J_\omega^{\perp}$, the nonlinear second-order planar Hall response $J_{2 \omega}^{\perp}$, and the nonreciprocal longitudinal response $J_{2 \omega}^{\|}$. $\theta$ and $\alpha$ denote the magnetic field angle and crystalline orientation angle relative to the electric field, respectively.}
    \label{fig:fig1}
\end{figure}

In this work, we show that an in-plane magnetic field provides precisely such a symmetry-resolving perturbation. For clarity, we retain only the Zeeman coupling to spin, as illustrated in Fig.~\ref{fig:fig1}. The field breaks the mirror and emergent $C_{2z}$ constraints while preserving a planar transport geometry, thereby converting otherwise hidden Berry curvature and quantum metric structures into measurable responses.

\begin{figure}
    \centering
    \includegraphics[width=1\linewidth]{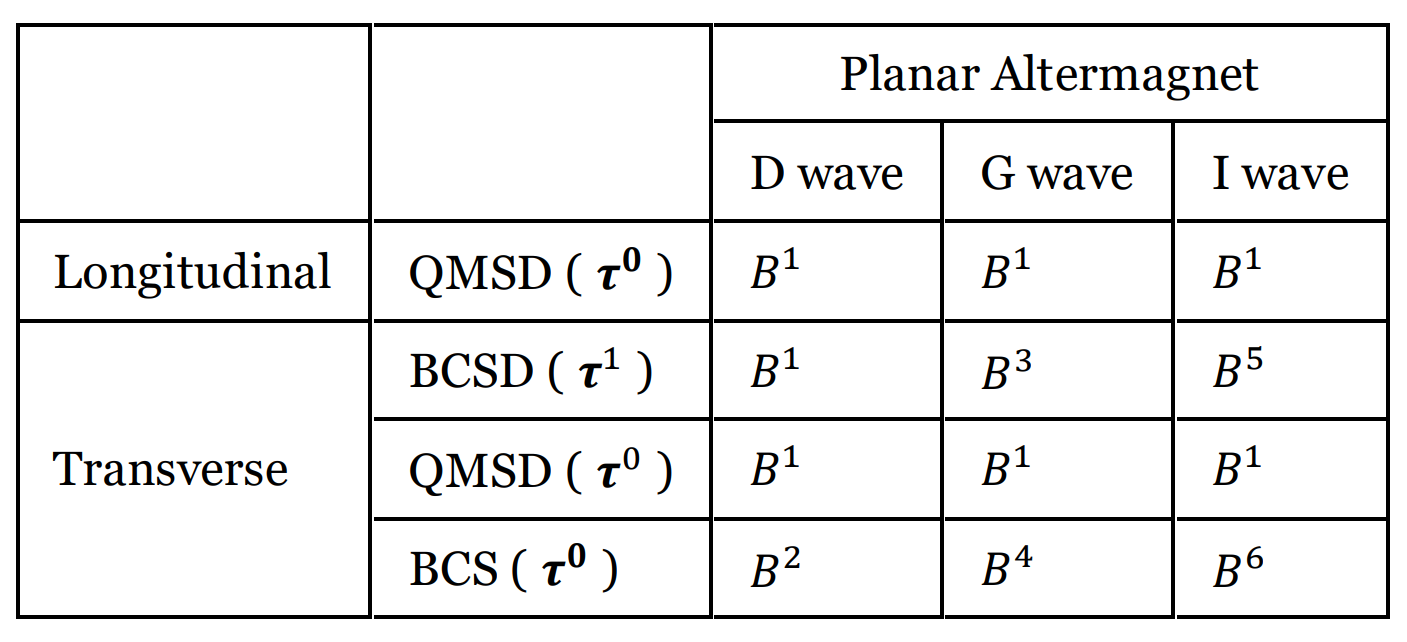}
    \caption{Quantum-geometric fingerprint table for planar altermagnets with different $C_n\mathcal{T}$ magnetic geometries under an in-plane magnetic field. The leading magnetic field powers and angular periodicities of the planar responses distinguish different altermagnetic order parameters.}
    \label{fig:fig2}
\end{figure}

The central observables are not only the magnitudes of the induced currents, but also their leading magnetic field powers and angular periodicities. These quantities are fixed by the $C_n\mathcal{T}$ magnetic symmetry and therefore serve as quantum-geometric fingerprints of the altermagnetic order. We identify three complementary fingerprint channels: the linear planar Hall response governed by Berry curvature susceptibility (BCS), the nonlinear planar Hall response governed by Berry curvature susceptibility dipoles (BCSD) and quantum metric contributions, and the nonreciprocal longitudinal response governed by the quantum metric susceptibility dipole (QMSD). As summarized in Fig.~\ref{fig:fig2}, this planar-magnetotransport scheme distinguishes $d$-, $g$-, and $i$-wave altermagnetic orders via symmetry-selected field scaling and angular harmonics.

%%%%%%%%
\emph{General theory of quantum-geometric fingerprints.}---A quantum-geometric fingerprint requires two ingredients: a geometric quantity that controls the transport response and a symmetry rule that imprints unique signatures onto the response, such as its leading scaling and angular dependence on an in-plane magnetic field. In altermagnets, the relevant geometric quantities are the Berry curvature and the band-normalized quantum metric. Because their zero-field contributions are constrained by mirror and emergent $C_{2z}$ symmetries, the leading magnetic field derivatives become the intrinsic objects that encode the altermagnetic order.

We start by sketching the derivation of the linear and nonlinear conductivities induced by quantum geometry, which take the forms $j^b=\sigma^{ab}E^a$ and $j^c=\sigma^{ab;c}E^aE^b$, where $j^a$ is the response current and $E^a$ is the external electric field. The quantum geometry of a Bloch Hamiltonian $H$ is described by the quantum metric and Berry curvature, the real symmetric and imaginary antisymmetric parts of the quantum geometric tensor~\cite{Jiang_2024,gao2025quantumgeometryphenomenacondensed,Kang2025}, $Q_n^{ab} = \sum_{m \neq n} A^a_{nm}A^b_{mn}=  g_n^{ab} - \frac{i}{2}  \Omega_n^{ab}.$ Here $A^b_{mn} = \langle u_m|\partial_b u_n\rangle $ is the interband Berry connection, and $| u_n\rangle$ is the $n$-th cell-periodic Bloch eigenstate of $H$.

The linear Hall conductivity is given by the integral of Berry curvature over occupied states~\cite{2010RvMP...82.1539N}, which reads
\begin{equation}
\sigma^{x y}=\frac{e^2}{\hbar} \int_{n\bm{k}} \Omega_n^{x y}f_n.
\end{equation}
The second-order nonlinear conductivity is given by 
\begin{equation}
\begin{split}
\sigma^{a b ; c}&=\frac{e^3 \tau}{\hbar^2} \int_{n\boldsymbol{k}} \frac{f_n}{2}\left(\partial_{k^a} \Omega_n^{b c}+\partial_{k^b} \Omega_n^{a c}\right)\\
&-\frac{e^3}{2\hbar} \int_{n\boldsymbol{k}}f_n\left[ \partial_{k^c} \tilde{g}_n^{a b}-2\left(\partial_{k^a} \tilde{g}_n^{b c}+\partial_{k^b} \tilde{g}_n^{a c}\right)\right],
\end{split}
\end{equation}
where the first term is contributed by the Berry curvature dipole \cite{PhysRevLett.115.216806,PhysRevB.107.115142, Lu2024} and the second term is contributed by the band-normalized quantum metric $\tilde{g}_n^{ab}=-\frac{\partial g^{ab}_n}{\partial\varepsilon_n} = \sum_{m\neq n} \frac{A^a_{nm}A^b_{mn} + A^b_{nm}A^a_{mn}}{\varepsilon_n - \varepsilon_m }$ \cite{10.1093/nsr/nwae334,https://doi.org/10.1002/advs.202514818}, which can be derived from the quantum kinetic theory~\cite{PhysRevB.108.L201405,Zhu2025}.

%%%%%%%%%%%%%%%%%%%%%%%%%%%%%%%%%%%%%%%%%%%%%%%%%%%%%%%%%%
%\begin{figure}
    %\centering
    %\includegraphics[width=1\linewidth]{QM.pdf}
%    \caption{Band-normalized quantum metric component multiplied by  group velocity $\tilde{g}_{xx}v_x$ of $d_{xy}$ planar altermagnet with: (a) $J=0.5 \mathrm{eV}\cdot nm^2$; (b) $J=0$. Other parameters: $v_\text{R}=1 \mathrm{eV}\cdot nm,\ g_s\mu_B B=0.2$ eV.}
    \label{fig:fig3}
%\end{figure}
%%%%%%%%%%%%%%%%%%%%%%%%%%%%%%%%%%%%%%%%%%%%%%%%%%%%%%%%%%

In an altermagnet, an in-plane magnetic field acts as a symmetry-resolving parameter. The total Hamiltonian is $H=H_0+H_B$, with $H_0$ describing the bare system and $H_B$ denoting the Zeeman energy $H_B=-\mathbf{h} \cdot \boldsymbol{\sigma}=-g_s\mu_BB\cos\theta\sigma_x-g_s\mu_BB\sin\theta\sigma_y$ induced by an in-plane magnetic field $B$~\cite{PhysRevB.82.045122}. Here $g_s$ is the Land\'{e}-g factor, $\mu_B$ is the Bohr magneton, and $\theta$ is the angle between the magnetic field and the current direction (x-axis).

We define the $m$-th order susceptibilities of the linear and nonlinear conductivities with respect to the magnetic field via the expansions $\sigma^{xy}=\kappa^{xy(m)} B^m$ and $\sigma^{ab;c}=\chi^{ab;c(m)}B^m$, respectively. Importantly, the leading nonzero orders of $\kappa$ and $\chi$ are not arbitrary perturbative coefficients: they are selected by the $C_n\mathcal{T}$ magnetic geometry. To identify their quantum geometric origin, we define the corresponding field susceptibilities of the Berry curvature and band-normalized quantum metric as
\begin{equation}
\alpha_n^{b c(m)}=\frac{1}{m!}\partial_B^m \Omega_n^{b c}|_{B \rightarrow 0},\label{eq:Berry_sus} \quad G_n^{b c(m)}=\frac{1}{m!}\partial_B^m \tilde{g}_n^{b c}|_{B \rightarrow 0}    
\end{equation}
where $\alpha_n^{bc(m)}$ and $G_n^{bc(m)}$ are the $m$-th order Berry curvature susceptibility (BCS) and the band-normalized quantum metric susceptibility (QMS) defined in the mixed space of crystal momentum $\bm{k}$ and the parameter $B$.

Here, we derive an explicit expression of the first order band-normalized QMS: ~\cite{2026PNAS..12306751H,cui2025electric,PhysRevLett.130.126303,PhysRevB.108.075155}
\begin{equation}
\begin{aligned}
G_n^{ab(1)}(\mathbf{k})=&2\mathrm{Re} \sum_{m \neq n}[  \frac{3\left(S_{m m}-S_{n n}\right) v_{n m}^a v_{m n}^b}{\left(\varepsilon_m-\varepsilon_n\right)^{4}} \\
& +\sum_{l \neq m} \frac{S_{l m} v_{n l}^a v_{m n}^b+S_{l m}^* v_{n m}^a v_{l n}^b}{\left(\varepsilon_m-\varepsilon_n\right)^3\left(\varepsilon_m-\varepsilon_l\right)} \\
&+\sum_{l \neq n} \frac{S_{l n}^* v_{l m}^a v_{m n}^b+S_{l n} v_{n m}^a v_{m l}^b}{\left(\varepsilon_m-\varepsilon_n\right)^3\left(\varepsilon_n-\varepsilon_l\right)}].\label{eq:metric_sus}
\end{aligned}
\end{equation}
Here $S_{nm}=-g_s \mu_B \left\langle u_n(\boldsymbol{k})\right|  (\cos(\theta)\sigma_x+\sin(\theta)\sigma_y)\left|u_m(\boldsymbol{k})\right\rangle$, and $v^{x(y)}_{n m}=\left\langle u_n(\boldsymbol{k})\right| \frac{\partial H}{\partial k_{{x(y)}}}\left|u_m(\boldsymbol{k})\right\rangle$ is the velocity matrix element.

\begin{figure}[t]
    \centering
    \includegraphics[width=\linewidth]{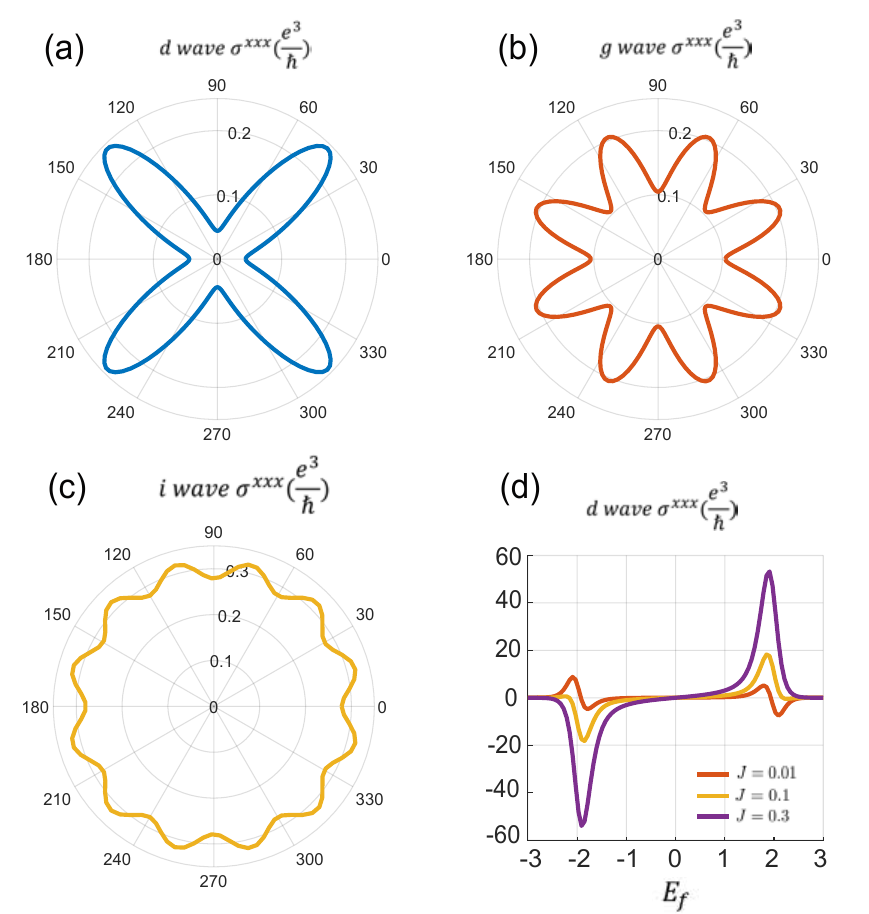}
    \caption{Nonreciprocal longitudinal conductivity $\sigma^{x;xx}=\chi^{x;xx}|B|$ for $d$-, $g$-, and $i$-wave altermagnets. Panels (a)--(c) show the $\alpha$ dependence of $\sigma^{x;xx}$ for different order parameter symmetries. Panel (d) shows the Fermi energy dependence of $\sigma^{x;xx}$. Parameters are $J_d=J_g=J_i=0.4$, $v_R=0.2$, $g_s\mu_BB=0.01$, $T=0.01$ and for (a)-(c) $\mu=0.05$ in units of $t_0$.}
    \label{fig3}
\end{figure}

In the following sections, we use this susceptibility framework to identify three fingerprint channels of $C_n\mathcal{T}$ magnetic geometry. We focus on intrinsic contributions and do not study extrinsic transport terms scaling as $\tau^n$ with $n>1$, which can be distinguished experimentally by their scattering-time dependence~\cite{Wang2023,Li2024}.

\emph{Quantum-metric fingerprints in nonreciprocal transport.}---We first identify the longitudinal fingerprint associated with the QMSD. To illustrate the effect and the associated geometric quantities, we consider an effective $k\cdot p$ Hamiltonian for a quasi-two-dimensional noncentrosymmetric altermagnetic metal with Rashba SOC, or equivalently a heterostructure composed of an altermagnet and a Rashba electron gas~\cite{PhysRevX.12.031042,Yang2026}. The model is
\begin{equation}
H_0=t_0k^2+\hat H_s-\mu ,
\label{eq:model}
\end{equation}
where $\mu$ is the chemical potential, $t_0k^2$ is the term of kinetic energy, and
\begin{equation}
\hat H_s=\Delta_{\bm{k}}(\alpha)\sigma_z
+v_\text{R}(\sigma_x k_y-\sigma_y k_x).
\label{eq:spinH}
\end{equation}
The term $\Delta_{\bm{k}}(\alpha)\sigma_z$ describes altermagnetic order, with $\alpha$ the angle between the crystalline axis and the current direction. We consider the representative altermagnetic orders
\begin{equation}
\Delta_{\bm{k}}(\alpha)=
\begin{cases}
J_d k'_x k'_y, & d\text{-wave},\\
J_g k'_x k'_y\left(k_x'^2-k_y'^2\right), & g\text{-wave},\\
J_i k'_x k'_y\left(k_x'^2-3k_y'^2\right)\left(3k_x'^2-k_y'^2\right), & i\text{-wave},
\end{cases}
\label{eq:orders}
\end{equation}
where $k'_x=k_x\cos\alpha+k_y\sin\alpha$ and $k'_y=k_y\cos\alpha-k_x\sin\alpha$. Candidate $d$-wave altermagnets include Rb-doped V$_2$Te$_2$O and K-doped V$_2$Se$_2$O~\cite{zhang2024crystalsymmetrypairedspinvalleylockinglayered,Jiang_2025}, while $g$- and $i$-wave orders may be realized in twisted magnetic van der Waals materials~\cite{PhysRevLett.133.206702}. In quasi-two-dimensional materials, an in-plane magnetic field produces a sizable Zeeman coupling while its orbital effect can be neglected. For instance, on the surface of Bi$_2$Se$_3$ with 5\% Cr doping, a weak in-plane field of approximately $0.2$~T yields a Zeeman energy of order $10$~meV~\cite{yrs7-m6zy,10.1063/5.0160335,PhysRevLett.130.036702}; the altermagnetic order remains robust because its spin splitting is much larger, on the order of $0.5$~eV~\cite{yrs7-m6zy}.

%And we restrict our investigation to the $\mathbf{J} \cdot \mathbf{B}$ -type MCA; other variants, such as the $\mathbf{J} \cdot(\mathbf{P} \times \mathbf{B})$ type typically found in polar systems, are not addressed in this article.   As a hallmark of nonreciprocal transport,
The nonreciprocal longitudinal response is the quantum-metric fingerprint of the altermagnetic order. Second-order nonlinear transport appears once the in-plane field breaks $C_{2z}$ symmetry. The nonlinear magnetotransport is described by $J^c=\sigma^{ab;c}(\bm{B})E_aE_b$. Under a global $C_{2z}$ rotation, both $\bm{J}$ and $\bm{E}$ change sign, so $\sigma^{ab;c}(-\bm{B})=-\sigma^{ab;c}(\bm{B})$. A $J\propto BE^2$ term is therefore symmetry allowed, producing magnetochiral anisotropy (MCA)~\cite{PhysRevLett.87.236602,PhysRevLett.117.146603,Jiang_2025MCA}. The total current may be written as
\begin{equation}
J=\sigma_0E+\chi BE^2\simeq
\left(\sigma_0+\frac{\chi BJ}{\sigma_0}\right)E ,
\end{equation}
which gives a bilinear correction to the longitudinal conductivity,
\begin{equation}
\sigma=\sigma_0+\frac{\chi^{xx;x(1)}}{\sigma_0}\,\bm{J}\cdot\bm{B}.
\label{eq:MCA}
\end{equation}
Here $\sigma_0$ is the ordinary Drude conductivity. We focus on the $\bm{J}\cdot\bm{B}$ type of MCA; other variants such as $\bm{J}\cdot(\bm{P}\times\bm{B})$ common in polar systems~\cite{7nxc-j62y} are absent here. In particular, the second-order longitudinal conductivity $\chi^{xx; x}$ is induced by the QMSD,
\begin{equation}
\chi^{xx;x(1)}
=\frac{3e^3}{2\hbar}\int_{n\bm{k}}
\left(f_n\partial_{k_x}G^{xx(1)}_n
+\partial_{k_x}\tilde g^{xx}_n\,\partial_B f_n\right).
\label{eq:chixxx}
\end{equation}

Notably, the QMD-induced MCA effect involves scattering processes rather than being a purely quantum mechanical perturbative result ~\cite{https://doi.org/10.1002/advs.202514818,tang2026absencequantummetricinducedintrinsiclongitudinal}.  However, as the prefactor $f(\tau)$ converges to a finite constant in the zero-frequency limit of the external field, we characterize this response as an intrinsic effect within the context of this study.
When the chemical potential approaches the Dirac point, $\mu=\pm v_\text{R} k_F$, the QMS contribution takes the form
\begin{equation}
\chi^{xx;x(1)}
=\frac{1}{v_\text{R}^2}\left(\frac{\mu}{v_\text{R}}\right)^{-2}
+\frac{\pi n^2J^2\sin^2(n\alpha)}{v_\text{R}^4}
\left(\frac{\mu}{v_\text{R}}\right)^{2n-4},
\label{eq:analyticQMSD}
\end{equation}
for $d_{xy}$-, $g_{xy(x^2-y^2)}$-, and $i_{xy(x^2-3y^2)(3x^2-y^2)}$-wave orders with $n=2,4,6$, respectively. Details are provided in the Supplemental Material (SM)~\cite{SM}.

\begin{figure*}
    \centering
    \includegraphics[width=0.9\linewidth]{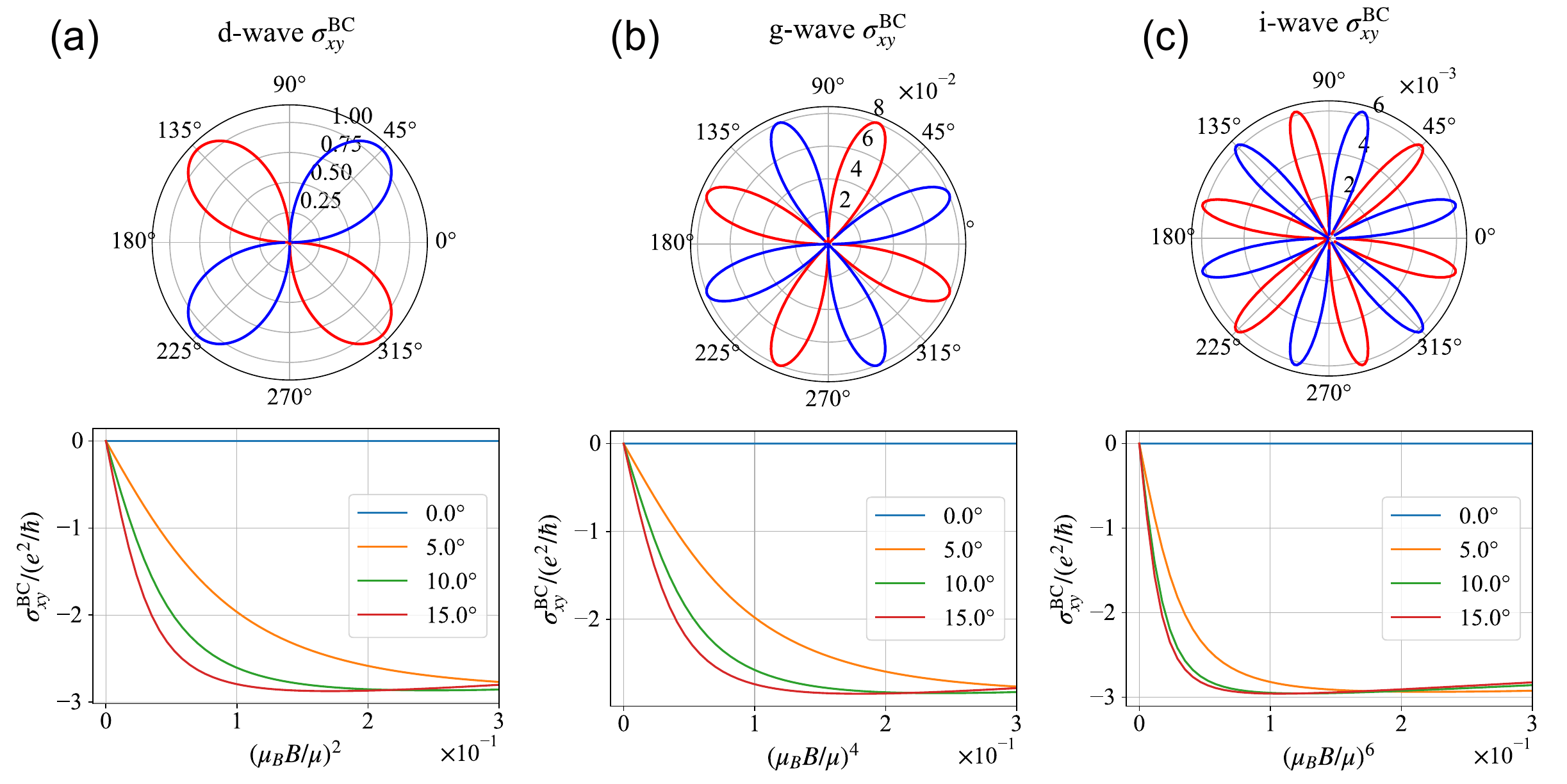}
    \caption{Linear Hall conductivity of (a) $d$-, (b) $g$-, and (c) $i$-wave altermagnets. Red and blue denote positive and negative values, respectively. Upper panels show the angular dependence on the magnetic field direction $\theta$. Lower panels show the magnetic field-magnitude dependence for selected field orientations. Parameters are $J_d=J_g=J_i=1$, $v_R=1$, and $T=0.1$ in units of $t_0$.}
    \label{fig:fig4}
\end{figure*}

The first term in Eq.~\eqref{eq:analyticQMSD} is an isotropic QMSD present even without altermagnetic order~\cite{doi:10.1126/science.adq3255}. The second term is an anisotropic QMSD contribution generated by altermagnetism. Because this leading altermagnetic term of $\chi^{xx;x(1)}$ is even under time reversal, it scales as $J^2$. Consequently, the $\alpha$ dependence of $\chi^{xx;x(1)}$ has period $\pi/n$, directly reflecting the $C_{2n}\mathcal{T}$ symmetry. This $2n$-fold angular structure of MCA provides a transport fingerprint of the underlying altermagnetic symmetry. Numerical calculations in Figs.~\ref{fig3}(a)--\ref{fig3}(c) confirm the predicted oscillation patterns: $d$-, $g$-, and $i$-wave orders exhibit four-, eight-, and twelvefold symmetries, respectively. Moreover, the altermagnetic order acts as an MCA amplifier, for $\tilde{g}_{xx}v_x$ is drastically enhanced (see SM~\cite{SM}). As shown in Fig.~\ref{fig3}(d), the resulting additional altermagnetic contribution in Eq.~\eqref{eq:analyticQMSD} substantially increases the second-order conductivity, suggesting a practical route to large MCA signals in nonmagnetic Rashba systems proximitized by altermagnetic layers.

The nonlinear planar Hall response, $j^y=\chi^{xx;y(1)}BE_x^2$, provides a transverse counterpart of this quantum-metric fingerprint and also receives a QMSD contribution,
\begin{equation}
\begin{split}
\chi^{aa;b(1)}
=&-\frac{e^3}{2\hbar}\int_{n\bm{k}}
\bigg[(\partial_{k_b}G^{aa(1)}_n-4\partial_{k_a}G^{ab(1)}_n)f_n\\
&+\partial_Bf_n
(\partial_{k_b}\tilde g^{aa}_n-4\partial_{k_a}\tilde g^{ab}_n)\bigg].
\end{split}
\label{eq:nonlinearHallQMSD}
\end{equation}
Its field scaling and angular periodicities in $\theta$ and $\alpha$ are analogous to those of the longitudinal response $\chi^{xx;x}$; a detailed analysis is given in the SM~\cite{SM}.

\emph{Berry curvature fingerprints in planar Hall responses.}---We next discuss Hall-type fingerprints associated with Berry curvature susceptibility. The planar magnetic field induced Hall conductivity is obtained by integrating the BCS defined in Eq.~\eqref{eq:Berry_sus}. Because $C_{2z}$ symmetry requires $\sigma^{xy}(-\bm{B})=\sigma^{xy}(\bm{B})$, the leading linear Hall response must be even in $B$, consistent with Onsager reciprocity for even-parity magnetic order. The Hall conductivity susceptibility is
\begin{equation}
\kappa^{xy(m)}
=\frac{e^2}{\hbar}\int_{n\bm{k}}
\left[\alpha^{xy(m)}_n f_n+(\partial_B^m f_n)\Omega^{xy}_n\right],
\label{eq:kappa}
\end{equation}
where $\alpha^{xy(m)}_n$ is the $m$th-order BCS.

Fig.~\ref{fig:fig4} shows the linear Hall conductivity $\sigma^{xy}(B,\theta)$ for $d_{xy}$-, $g_{xy(x^2-y^2)}$-, and $i_{xy(x^2-3y^2)(3x^2-y^2)}$-wave altermagnets. Each order exhibits a distinct angular periodicity, magnetic field scaling, and number of zeros. The lowest nonvanishing $\alpha^{xy(m)}_n$ occurs at $m=2,4,6$ for the $d$-, $g$-, and $i$-wave orders, respectively; the corresponding susceptibilities $\kappa^{xy(m)}$ can be extracted from the small-$B$ slopes in Fig.~\ref{fig:fig4}. The pair of observables--the leading field power and the field-angle harmonic--therefore provides a linear Hall fingerprint for identifying the symmetry of the altermagnetic order.

\begin{figure*}[t]
    \centering
    \includegraphics[width=0.9\linewidth]{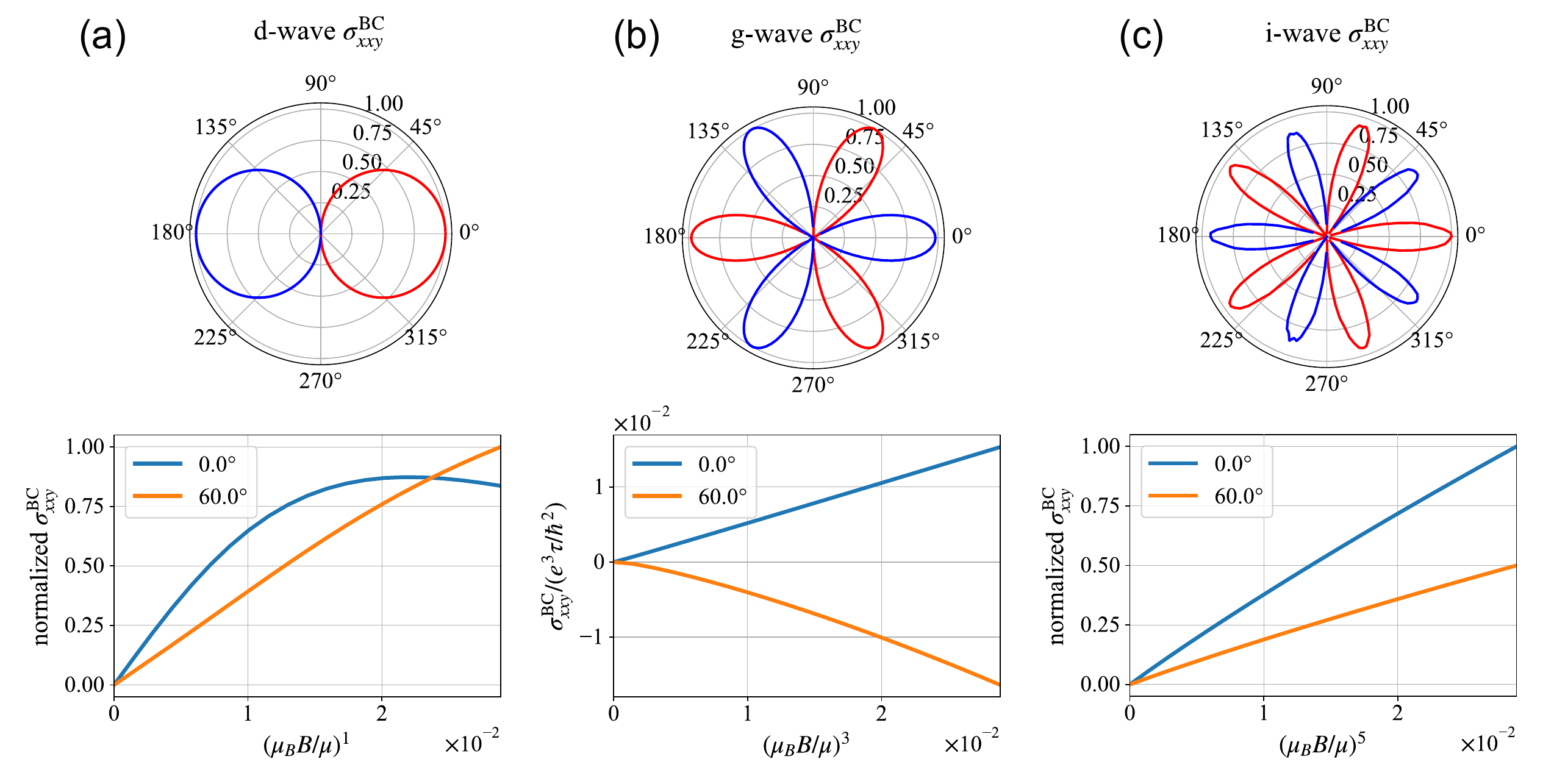}
    \caption{BCSD-induced second-order Hall conductivity $\sigma^{\rm BC}_{xx;y}$ for (a) $d$-, (b) $g$-, and (c) $i$-wave altermagnets. Red and blue denote positive and negative values, respectively. Upper panels show the angular dependence on the magnetic field direction $\theta$ at $\mu_BB/\mu=0.004$ in arbitrary units. Lower panels how the magnetic field-magnitude dependence at selected field orientations. $\sigma^{\rm BCSD}_{xx;y}$ is plotted in units of $e^3\tau/\hbar^2$. Parameters are $J_d=J_g=J_i=1$, $v_R=1$, and $T=0.1$ in units of $\mu$.}
    \label{fig:fig5}
\end{figure*}
These results follow from symmetry. The Hall conductivity $\sigma^{xy}$ is a pseudoscalar and is therefore invariant under a global spatial rotation of the entire system, including the magnetic field. For a $C_{2n}\mathcal{T}$ altermagnet, $\sigma^{xy}$ is $n$-fold symmetric in the field orientation, so the lowest allowed term scales as $\sigma^{xy}\propto B^n\sin(n\theta)$. This accounts for the $B^n$ scaling. The zeros originate from mirror symmetry: at zero field, mirror symmetry enforces $\sigma^{xy}=0$. A generic in-plane field breaks this mirror symmetry and permits a finite planar Hall response. However, when the field is exactly perpendicular to a mirror plane, that mirror symmetry is preserved and the Hall response vanishes. Together with the $2\pi/n$ rotational periodicity, this produces exactly $2n$ zeros over a full $2\pi$ rotation.

The second-order Hall conductivity $\sigma^{xx;y}$ is odd in $B$. The higher-order BCSD contribution is
\begin{equation}
\chi^{aa;b(m)}
=\frac{e^3\tau}{\hbar^2}\int_{n\bm{k}}
\left[\partial_{k_a}\alpha^{ab(m)}_n f_n
+(\partial_B^m f_n)\partial_{k_a}\Omega^{ab}_n\right].
\label{eq:BCSD}
\end{equation}
The numerical results for the BCSD-induced nonlinear planar Hall conductivity, $\sigma^{xx;y(m)}=\chi^{xx;y(m)}|B|^m$, are shown in Fig.~\ref{fig:fig5}. The field scaling again depends on the altermagnetic order: $m=1,3,5$ for the $d$-, $g$-, and $i$-wave cases, respectively. These scalings provide a nonlinear Hall fingerprint complementary to the linear BCS response.

The $(n-1)$-fold rotational symmetry of $\sigma^{\rm BCSD}_{xx;y}$ in a $C_{2n}\mathcal{T}$ altermagnet follows from the angular structure of
$\sigma^{\rm BCSD}_{ab;c}\propto\int d^2k\,f\,\partial_{k_a}\Omega$. In the absence of a magnetic field, the Berry curvature has $n$-fold symmetry in momentum space, with the lowest nonconstant harmonic proportional to $\cos(n\alpha)$. The momentum derivative $\partial_{k_a}$ introduces a first harmonic $\sim\cos\alpha$, while the in-plane field couples linearly to momentum as $kB\cos(\alpha-\theta)$ and also carries a first harmonic. For the angular integral to be nonzero, the integrand must be independent of $\alpha$. Canceling the $n$th harmonic therefore requires $B^{n-1}$, yielding both the $B^{n-1}$ scaling and the $2\pi/(n-1)$ angular periodicity in $\theta$; a detailed derivation is provided in the Supplemental Material~\cite{SM}.

\emph{Conclusion and discussion.}---We have established planar magnetotransport as a quantum-geometric fingerprinting scheme for altermagnetic order. In two-dimensional altermagnets, an in-plane Zeeman field breaks the mirror and emergent $C_{2z}$ constraints that otherwise hide intrinsic Hall and second-order responses. This exposes the magnetic field susceptibilities of the Berry curvature and quantum metric, whose leading powers and angular periodicities are fixed by the underlying $C_n\mathcal{T}$ magnetic geometry.

The resulting fingerprints appear in three complementary transport channels. The nonreciprocal longitudinal response probes the QMSD and captures the crystalline orientation dependence of the altermagnetic order. The linear planar Hall response probes the BCS and displays order-specific magnetic field scaling. The nonlinear planar Hall response probes BCSD and QMSD contributions, providing a complementary second-order fingerprint. For representative $d$-, $g$-, and $i$-wave altermagnets, these channels exhibit distinct field power laws and angular harmonics, enabling symmetry selective identification of the altermagnetic order.

Based on the estimates shown in Figs.~\ref{fig3}--\ref{fig:fig5} and adopting the lattice constant $a\simeq4\times10^{-9}\,\mathrm{m}$ and scattering time $\tau\simeq10^{-13}\,\mathrm{s}$, we obtain $\sigma^{\rm QMSD}_{xx;x}\simeq30\,\mathrm{mA\,\mu m/V^2}$ for MCA, $\sigma^{\rm BCS}_{x;y}\simeq3.65\times10^{-3}\,\mathrm{mA/V}$ for the linear Hall effect, and $\sigma^{\rm BCSD}_{xx;y}\simeq148\,\mathrm{mA\,\mu m/V^2}$ for the BCSD-induced nonlinear Hall effect. These values are large enough to be experimentally resolved~\cite{doi:10.1126/science.adq3255,Wang2023}. Our results therefore show that planar magnetotransport can serve not only as a probe of band quantum geometry, but also as a practical diagnostic of magnetic geometry in altermagnetic materials.

\emph{Acknowledgments.}---This work was supported by the Ministry of Science and Technology, China, The New Cornerstone Foundation, the State Key Laboratory of Quantum Information Technologies and Materials, and the Hong Kong Research Grants Council through Grants No. MOST23SC01-A, No. RFS2021-6S03, No. C6053-23G, No. AoE/P-701/20, AoE/P-604/25R, No. 16309223, No. 16311424, and No. 16300325.

\bibliographystyle{apsrev4-1} 
\bibliography{References}

\end{document}